# A Method for Plasma Electron Temperature Evaluation for GOL-NB facility

**S. Polosatkin,**[a,b,c]

[a] *Budker Institute of Nuclear Physics,*
*630090, Lavrent'eva av.11, Novosibirsk, Russia*
[b] *Novosibirsk State Technical University,*
*630073, Prospekt K. Marksa 20, Novosibirsk, Russia*
[c] *Novosibirsk State University,*
*630090, Pirogova str. 1, Novosibirsk, Russia*

*E-mail*: s.v.polosatkin@inp.nsk.su

ABSTRACT: A new method for measuring the electron temperature of the plasma in GOL-NB facility is proposed. The proposed method is based on measuring the ratio of intensities of spectral lines emitted by fast atoms injected into the plasma. The beams of fast hydrogen atoms used for plasma heating or diagnostics contain atoms with full energy as well as atoms with fractional energies (E/2, E/3, E/18), which arise from the dissociation of molecular ions $H^+_2$, $H^+_3$, $H_2O^+$. Spectral lines of atoms with different energies (especially Ha) can be resolved due to Doppler shift caused by differences in atom velocities. For low-energy atoms, excitation occurs due to collisions with thermal electrons, while for high-energy atoms, collision processes with plasma ions are essential. Therefore, the intensity ratio of line fractions with different energies depends on the electronic temperature of the plasma and can be used for its measurement. At a beam energy of 24 keV, the method can be used to measure electronic temperatures up to 40 eV, which is of interest for experiments on the GOL-NB facility. To measure the temperature with an accuracy of 20 eV, it is necessary to measure the intensity ratio of lines with percentage precision and also measure the same precision of attenuation of the neutral beam passing through the plasma.

## Contents

## 1. Introduction

The GOL-NB facility [1] is a linear axisymmetric open trap with multimiror end plugs for plasma confinement. The main objective of experiments conducted in this facility is to investigate plasma confinement in a multimirror (periodically modulated along the axis) magnetic field. The GOL-NB facility consists of a 2.4 m long central cell with a central field strength of B = 0.3 T, and attached sections with strong fields, each approximately 3 m long with B = 4.5 T. Layout of the facility is shown in Figure 1. Plasma in the facility is created by a plasma gun located in the expander at one end of the facility and is heated using injection of beams of fast hydrogen atoms (neutral beams). Two neutral beam injectors (NBI) [2] with a total power of 1.1 MW inject fast neutrals with an energy of 24 keV into the plasma across the magnetic field of the facility at coordinates z = ±0.4 m (the longitudinal coordinate z is measured from the center plane of the trap). Passing through the plasma, the beams of fast atoms are partially ionized due to collisions with plasma electrons and ions, and the resulting fast ions are captured by the magnetic field of the facility. The captured ions oscillate along the axis of the facility between stopping points (coinciding with injection points) and gradually transfer their energy to the plasma electrons through elastic collisions.

The essence of the planned experiments on the facility is to compare the efficiency of plasma confinement in configurations with homogeneous (B = 4.5 T) and multimirror (B = 3.2/4.8 T, period of modulation 22 cm) fields in the end sections. The multimirror magnetic field should reduce longitudinal plasma losses and lead to an increase in plasma electron temperature (while maintaining a constant plasma density by proper plasma fueling in the central section). Calculations predict an increase in plasma temperature from 30-40 eV to ~100 eV in the most optimistic scenario of the experiment [3].

Currently, a series of experiments with a homogeneous configuration of the magnetic field has been conducted. In these experiments, a Langmuir probe [4], located at coordinate z = 87 cm, was used to measure the electron temperature. The temperature of the initial plasma created by the plasma gun was 5-10 eV, and during the injection of beams, the plasma temperature according to the probe readings increased to ~15 eV. It should be noted that the Langmuir probe is incompatible with fast ions in the plasma, so it was located at a significant distance from the region of motion of captured ions and plasma heating (z = ±40 cm). In addition, introducing a probe into the plasma can lead to significant cooling, so there are grounds to assume that the true plasma temperature in the central section is higher than the temperature measured by the Langmuir probe.

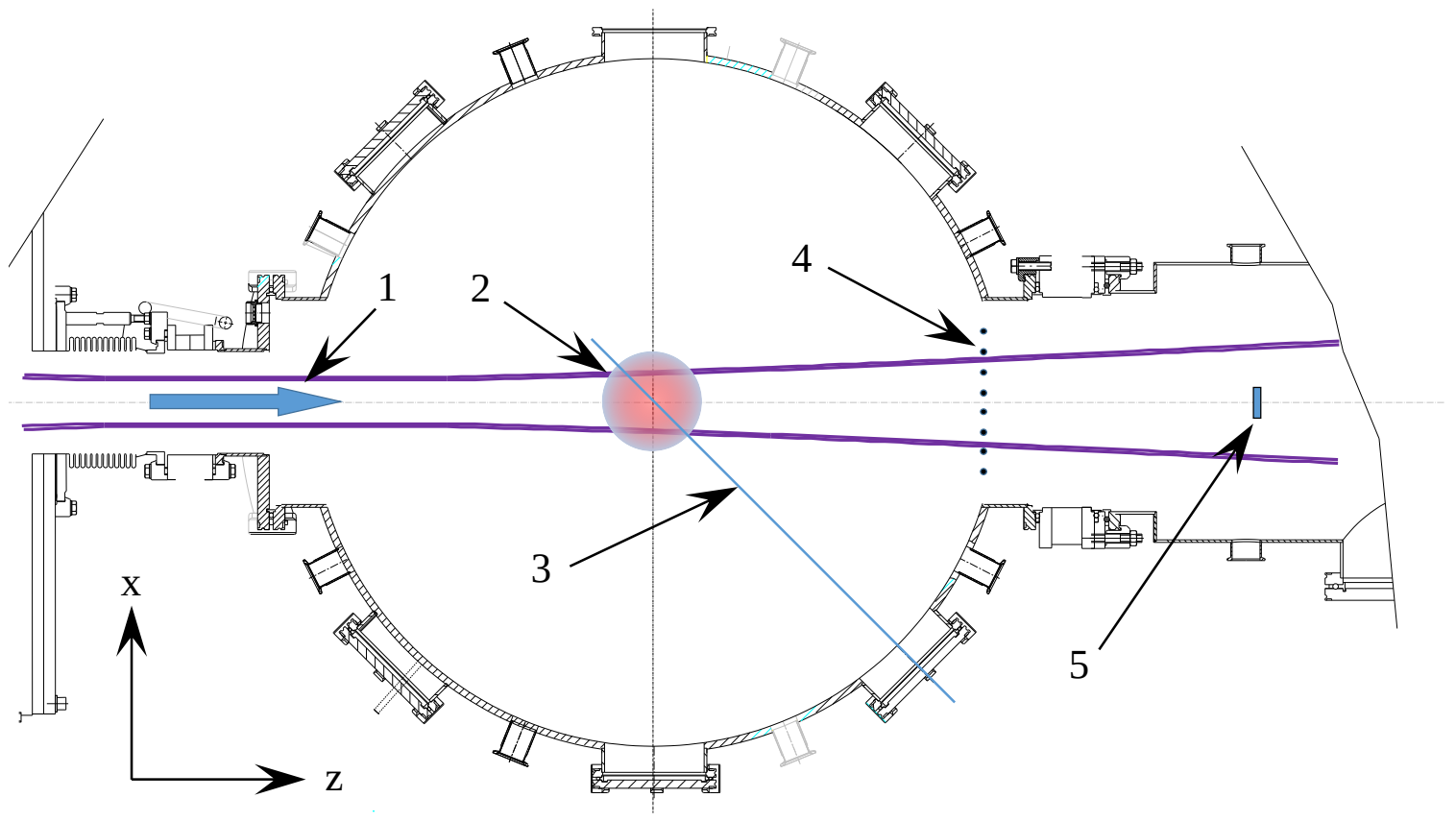

**Figure 1.** GOL-NB facility layout in the cross-section of a neutral beam injection. 1 – neutral beam, 2 – plasma, 3 – observation line of Doppler-shifted $H_\alpha$ radiation, 4 – beam profilometer, 5 – secondary emission beam attenuation probe

Therefore, the development of non-contact diagnostics for measuring the electron temperature of the plasma in the central section is a key challenge for the conducted and planned experiments. Currently, work is underway to create a Thomson scattering system on the GOL-NB facility. In this article, we propose another simple method for determining the electron temperature in the temperature range of 0-30 eV, based on measuring the intensity ratio of spectral lines of fast atoms injected by the neutral injection system.

## 2. A method of electron temperature measurements

The proposed method for electron temperature evaluation is based on the observation of hydrogen Hα spectral lines emitted by fast atoms injected into plasma. A feature of neutral beam injectors used for plasma heating is the presence of atoms with fractional energies in the neutral beam. Such beams contain aside from fast atoms with full energy equal to the accelerating voltage of the ion source also the atoms with fractional energies (E/2, E/3, E/18) that appear in the beam due to acceleration and follow dissociation of molecular ions $H^+_2$, $H^+_3$, $H_2O^+$.

Collisions between fast atoms with gas or plasma particles can lead to the excitation of atoms and subsequent emission of photons at the optical transitions of hydrogen atoms, such as the Balmer $H_\alpha$ line. It is important to note that when detecting radiation at an angle to the direction of beam motion, spectral lines of beam species with different energies can be separated from each other and from the background radiation due to the Doppler shift of the corresponding spectral lines. Doppler shift spectroscopy when the beam passes through a gas target is a standard method for determining the fractional composition of the beam [5,6].

In a plasma, excitation occurs due to collisions with electrons and ions. At low beam energy, excitation occurs only in electron collisions and strongly depends on electron temperature, while at high beam energy, ion collisions dominate, and the excitation rate becomes temperature independent. As a result, excitation rates for different beam fractions have different temperature dependencies. Calculated excitation rates for $H_\alpha$ in plasma for different atom energies are shown in Figure 2. Sources of atomic data for these calculations and their consistency are discussed in the next section. Bold lines in figure 1 correspond to the energies of beam species for conditions of the GOL-NB injection system (1.33 / 8 / 12 / 24 keV). As can be seen from the figure, the excitation reaction rates for atoms with different energies have different dependencies on the temperature of electrons. Therefore, the intensity ratio of lines corresponding to different beam fractions can be used to determine the electron temperature of the plasma.

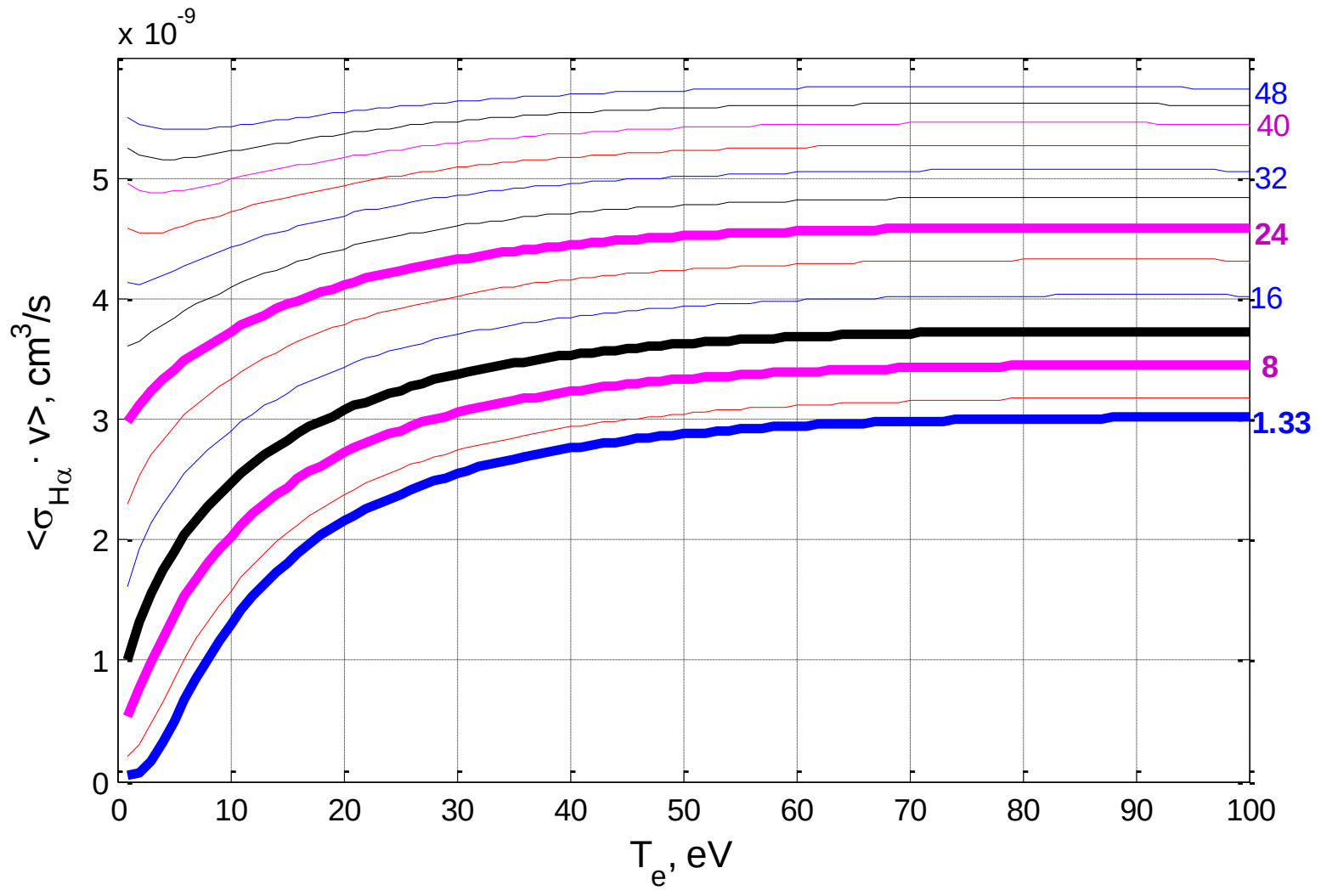

**Figure 2.** Doppler-shifted $H_\alpha$ excitation rates in plasma for different energies of a neutral beam. The beam energies are indicated near the right axis of the figure. Bold lines correspond to fractions of 24 keV neutral beam. The source and consistency of atomic data used in the calculations are discussed in section 3

Measurement of electron temperature based on the intensity ratio of Doppler-shifted $H_\alpha$ lines requires knowledge of the equivalent currents of different beam fractions. The most convenient and direct method for determining the fractional composition is by measuring the intensities of Doppler-shifted lines when the beam passes through a gas. These measurements can be made in the same geometry and with the same spectral apparatus as the measurement of line ratios in the plasma. For this, gas is injected into the chamber before beam injection in a calibration shot, but no plasma discharge is ignited.

Normalizing the intensities of lines when the beam passes through the plasma to the intensities of corresponding lines when the beam passes through gas allows for the exclusion of the influence of beam fractional composition, the sensitivity of radiation detection channels, as well as plasma and neutral gas concentrations.

Figure 3a shows calculated ratios of normalized intensities of beam fractions for the GOL-NB neural beam (E=24 keV), and Figure 3b – accuracies of measurement of the intensity ratios required for evaluation of electron temperature with accuracy 10%.

As can be seen from the figures, in the temperature range of up to 40 eV, which is of primary interest for current experiments on the GOL-NB, the intensity ratio has some temperature dependence and can be used for temperature determination. The ratio between lines with total energy E and energy E/18 has the strongest dependence. However, since atoms with energy E/18 are formed from dissociated water ions, the fraction of this component in the beam may vary from shot to shot, leading to temperature measurement errors. Additionally, the E/18 line is located close to the unshifted Hα line, which limits the accuracy of its amplitude measurement. Therefore, it is believed that the most promising approach is to use the ratio of lines with energies E and E/3.

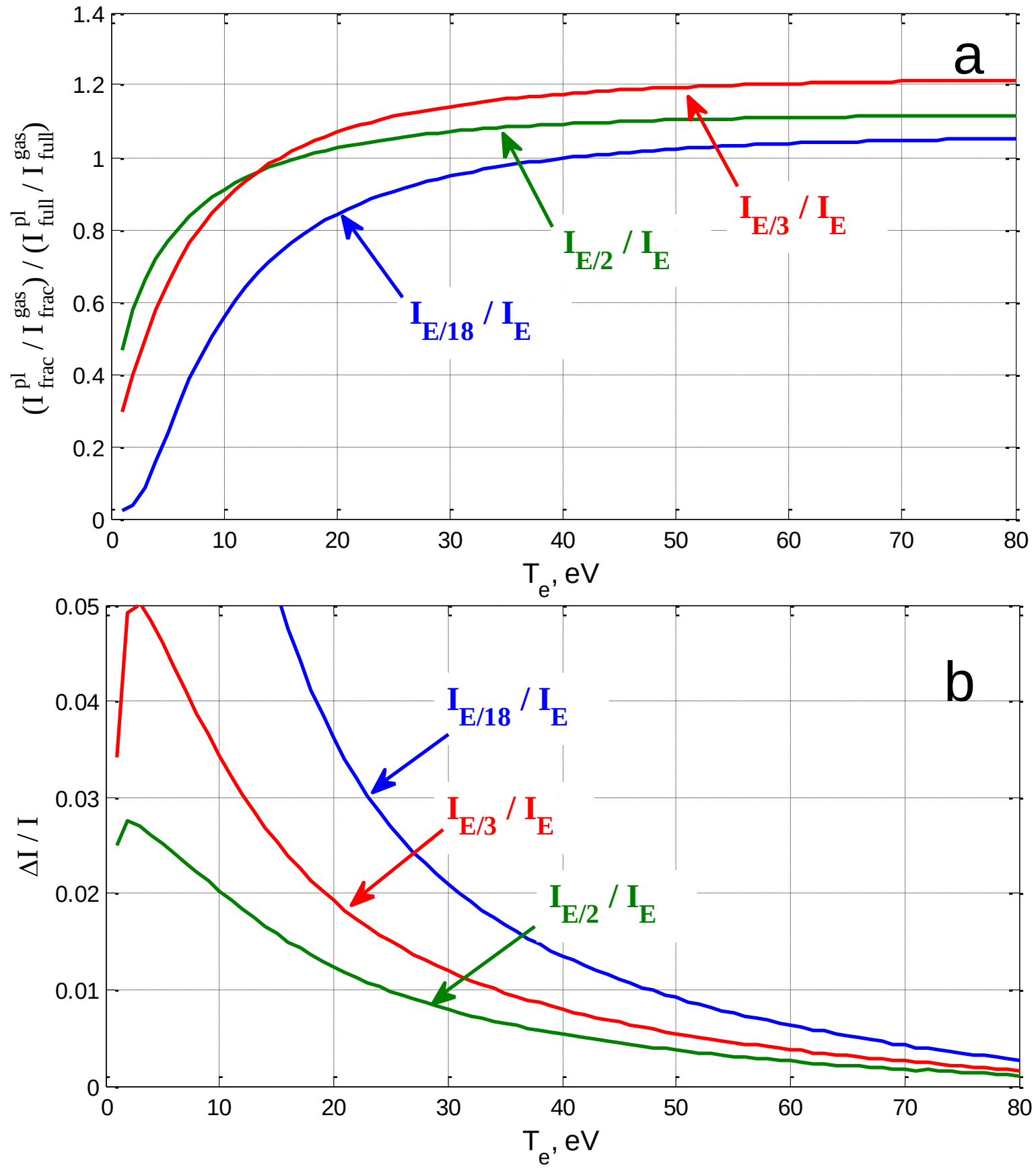

**Figure 3.** a) Ratios of normalized intensities of Doppler-shifted Hα for 24 keV beam fractions vs. electron temperature. b) Required accuracy of the intensity ratios measurements for evaluation of electron temperature with an accuracy of 10%.

## 3. Atomic data consistency

### 3.1 Sources of atomic data

Since the proposed method requires an accuracy of a few percent in the measurement of line intensities to determine the electron temperature, knowledge of excitation rates with the same high accuracy is necessary. For our calculations, we used excitation cross-sections of atoms in plasma from [7] and excitation cross-sections of neutrals colliding with hydrogen molecules from [8]. All datasets were taken from the IAEA Aladdin service [9]. To verify our calculations, we compared the results with publicly available OPEN ADAS data [10,11]. In OPEN ADAS, calculated hydrogen spectral lines electron impact photon emissivity rates vs. plasma temperature and density are available (for motionless hydrogen atoms) (files pec12#h_pju#h0.dat and pec96#h_pjr#h0.dat from the ADF15 library), as well as effective $H_\alpha$ beam emission rates vs. beam energy and plasma density for plasma temperature of 2 keV (file bme10#h_h1.dat).

### 3.2 $H_\alpha$ excitation model

The main uncertainty in the calculations is the choice of the sublevel population model with different orbital quantum numbers. The effects of population mixing on determining the fractional composition of atomic beams by Doppler-shift spectroscopy data are discussed in articles [12,13]. Here we suggest strong mixing of sublevel populations (TE - thermal equilibrium model). The reason is that integration of the cross-section of 3s-3p transition [14] over Maxvellian distribution gives a transition rate above $2.3\times10^{-5}$ $cm^3/s$ for temperature up to 50 eV, which corresponds to sublevel lifetime 4.2 ns even for plasma density $10^{13}$ $cm^{-3}$ that is suffisiently below radiative lifetime of 3s level (158 ns). It is easy to check that ADAS data are also calculated in the assumption of thermal equilibrium of sublevel populations even for the least tabulated density $5\times10^7$ $cm^{-3}$. For example, the ratio of photon emissivity rates of Balmer-α and Layman-β lines in ADAS data is exactly equal to the value of 0.79 – branching factor in the TE model. It should be noted that such strong mixing and fast decay of the 3s state means that there is no spatial transfer of excitation along with the movement of fast atoms, which could lead to errors in calculating intensity due to plasma non-uniformity across the radius.

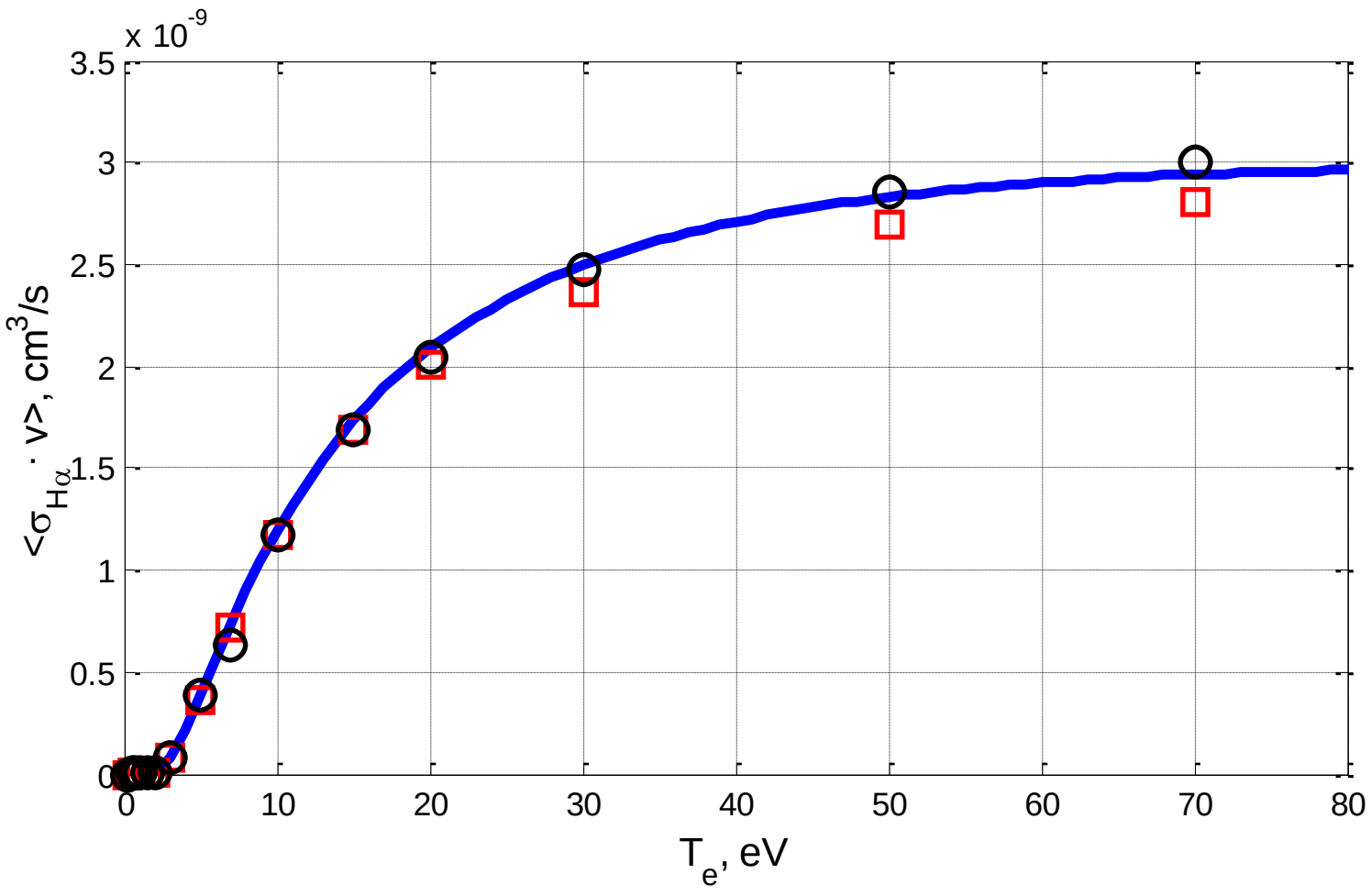

**Figure 4.** Electron impact $H_\alpha$ emissivity rates for motionless hydrogen atoms; solid line – our calculations, black rounds – state-resolved ADAS data (pec96#h_pjr), red squares - state-unresolved ADAS data (pec12#h_pju). ADAS data are taken for the least tabulated density $5\times10^7$ $cm^{-3}$.

In our model, we take in account only excitation from ground state to the levels n=3 and n=4 and radiative decay of these levels with branching factors 0.442 for $3\rightarrow2$ transition and 0.317 for $4\rightarrow3$ transition. State-unresolved excitation cross-sections $1s\rightarrow n=3$ and $1s\rightarrow n=4$ were used.

### 3.3 Comparison of atomic data

A comparison of the calculated electron impact $H_\alpha$ emissivity rate with OPEN ADAS data is shown in Figure 4. It should be noted that here we compared the emissivity rate calculations with available data from ADAS for motionless atoms (i.e., in the case of a spherically symmetric

Maxwellian distribution of bombarding electron atoms). To determine the emission rate for fast atoms, the excitation rate was averaged over a Maxwellian distribution shifted by the velocity of the fast atom.

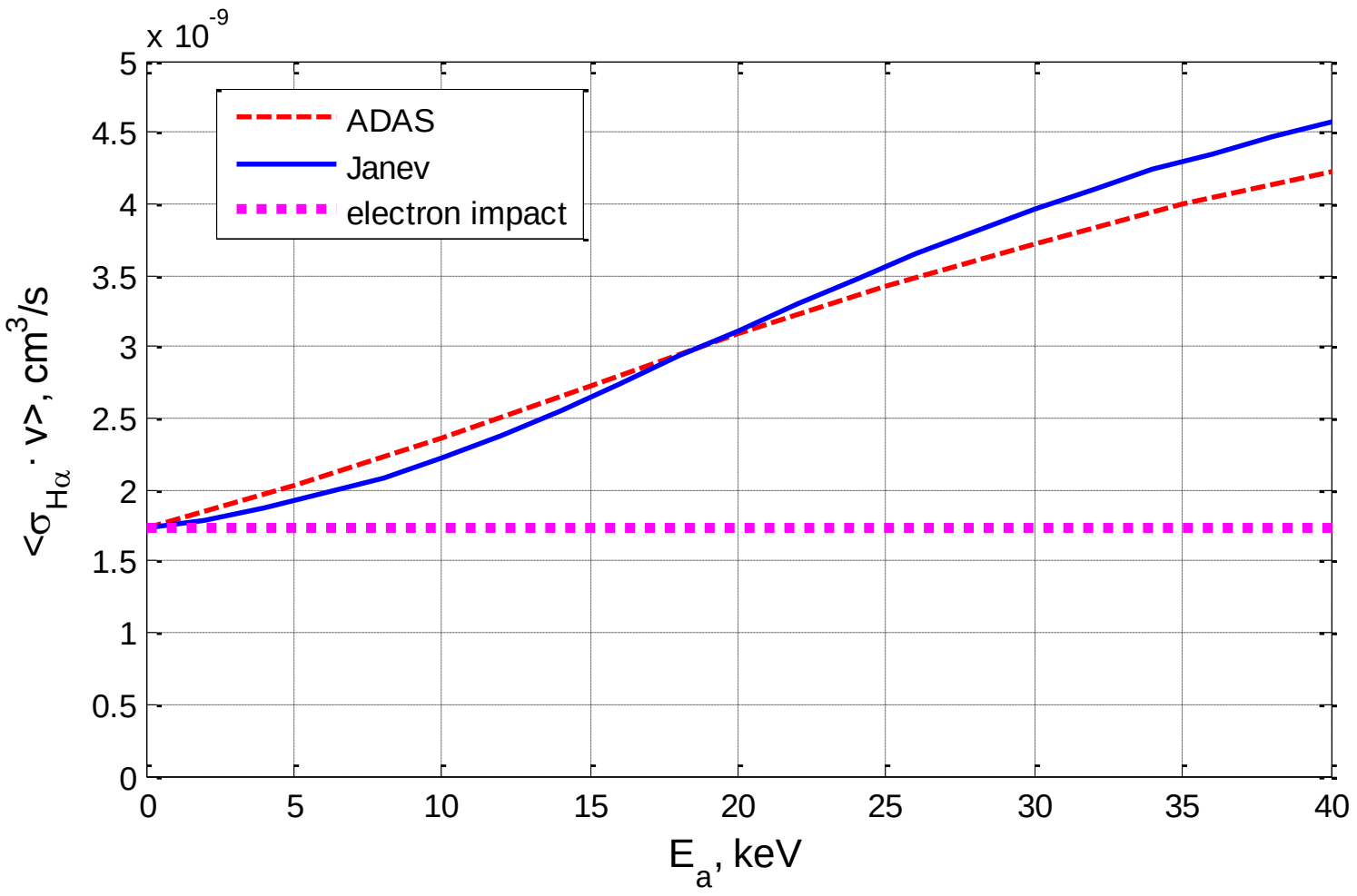

**Figure 5.** Net $H_\alpha$ beam emissivity rates in the plasma with $T_e$=2 keV; blue solid line- calculation using Janev cross-sections [7], red dash line – OPEN ADAS data [11], magenta dot line – electron impact emissivity rate. ADAS data are taken for the least tabulated density $5\times10^7$ cm$^{-3}$.

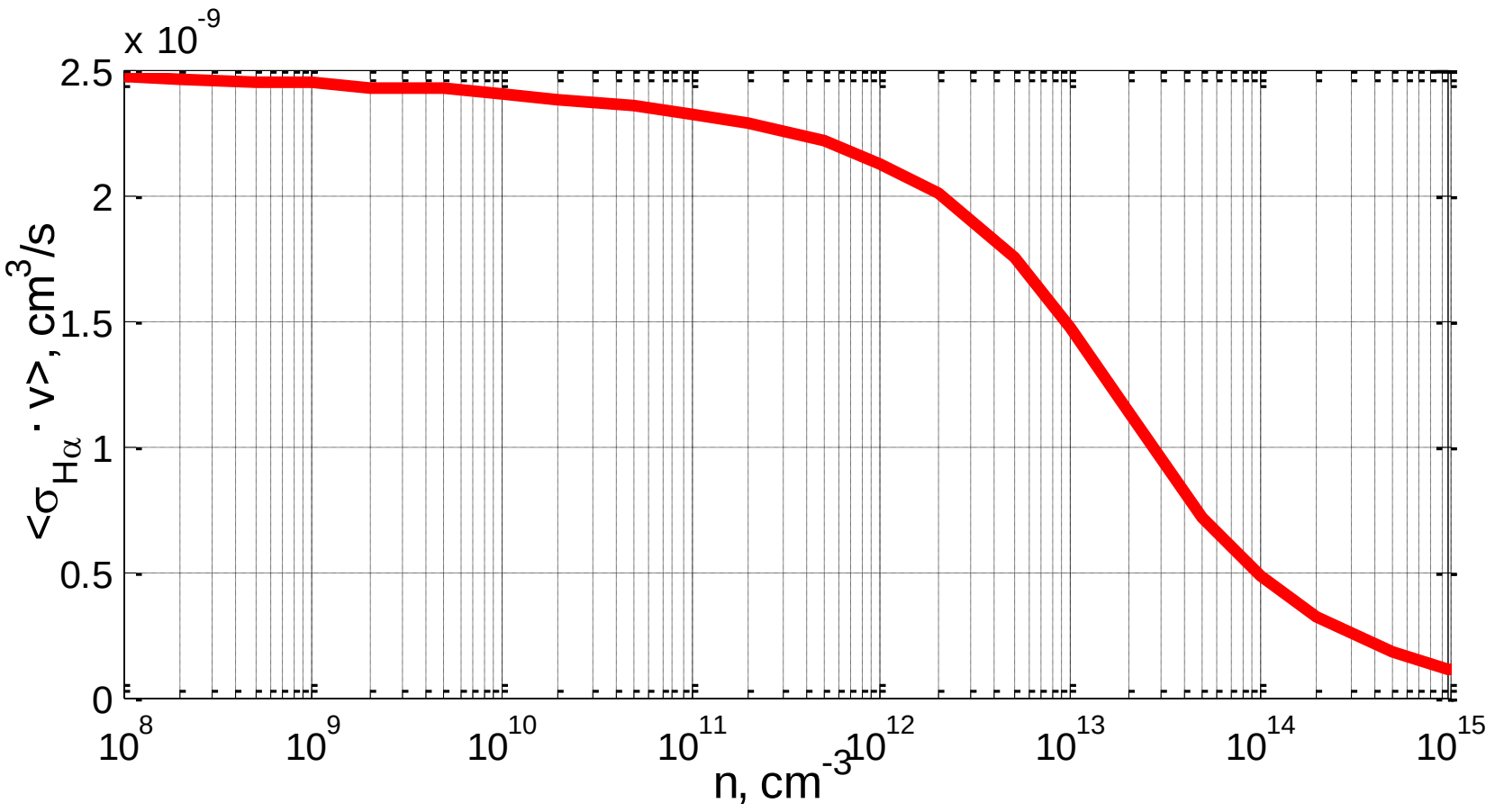

**Figure 6.** Density dependence of electron impact $H_\alpha$ emissivity rate for plasma temperature Te=30 eV

Verification of ion impact $H_\alpha$ emissivity rate was done by comparison of calculated beam emissivity rate in plasma with Te=2 keV with corresponding OPEN ADAS dataset (file bme10#h_h1.dat) (Figure 5). The blue solid line corresponds to calculations based on Janev cross-sections, red dashed line – to the tabulated OPEN ADAS dataset. The electron impact emissivity rate is shown by a magenta dot line. This means that for this temperature, the thermal velocity is much greater than the velocity of fast atoms, so this rate is practically independent of the beam energy.

Substantial (several percents) discrepancy in the emissivity rates was observed. Moreover there is the same discrepancy between datasets of Janev [7] and Barnett [8] cross-sections. So we decide to use the ion impact emissivity rate extracted from the OPEN ADAS dataset as a difference between net and electron impact emissivity rates.

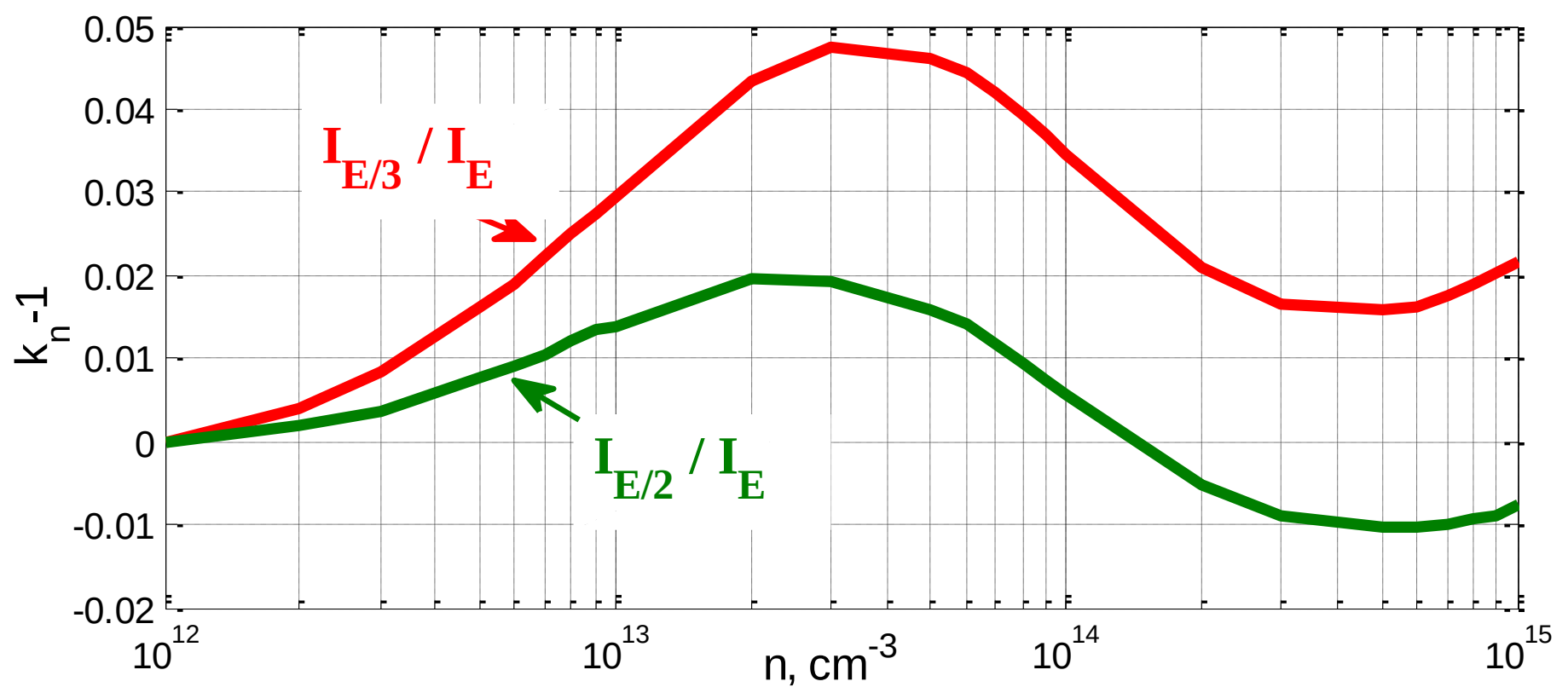

**Figure 7.** Density correction factors minus unity for half (green) and one-third (red) energy beam fractions

### 3.4 Density correction

Our model does not include collisional de-excitation which can lead to an appearance of dependence of emissivity from plasma density. ADAS predict a significant decrease of emissivities in the density range of interest of $10^{13}$-$10^{14}$ $cm^{-3}$ (see Figure 6). Fortunately, the dependence of emissivity rate on density for different energies of fast atoms is self-similar, and as a result, the ratio of these rates used to determine electron temperature only weakly depends on density. Moreover  this dependence can be accounted by using a calculated density correction factor defined as:

$$k_n = \frac{S_n^{E_frac} / S_{n0}^{E_frac}}{S_n^{E0} / S_{n0}^{E0}}$$

where $S_n^E$ – emissivity rate for energy of fast atom E in plasma with density n. The indices E_frac and E0 correspond to atoms with the fractional and full energy of a neutral beam, and indices n and n0 correspond to the plasma density and the density for which the emissivity rates are calculated. The plasma density for calculating the correction factor can be found by measuring the beam attenuation with a secondary emission profilometer, which is standard diagnostics at the GOL-NB. Calculated from the ADAS dataset density correction factors for half and one third energy beam fractions are shown in Figure 7. Neglecting these correction factors results in errors in intensity predictions of several percent, which does not meet the accuracy requirements for temperature measurements.

## 4. Beam attenuation effects

The proposed method for determining plasma temperature is based on the use of heating neutral beams, which must be captured by the plasma for effective heating. In current experiments on the GOL-NB facility, the attenuation coefficient of neutral beams due to capture in the plasma reaches 50%. Capture and attenuation of neutral beams occur due to charge exchange, as well as ionization during collisions of fast atoms with ions and electrons in the plasma. The effective cross-sections of these processes depend on the energy of fast atoms, so the attenuation coefficients differ for different beam fractions. This leads to the fact that in the region of registration of Doppler-shifted lines, the fractional composition of the beam differs from the initial composition before passing through the plasma, which must be taken into account when determining the plasma temperature. In addition, radiation registration is carried out not at a local point in space, but along a chord where the density of the plasma, the beam current density, and its fractional composition change. These effects must be taken into account in a model predicting the dependence of the intensity ratio of lines on temperature.

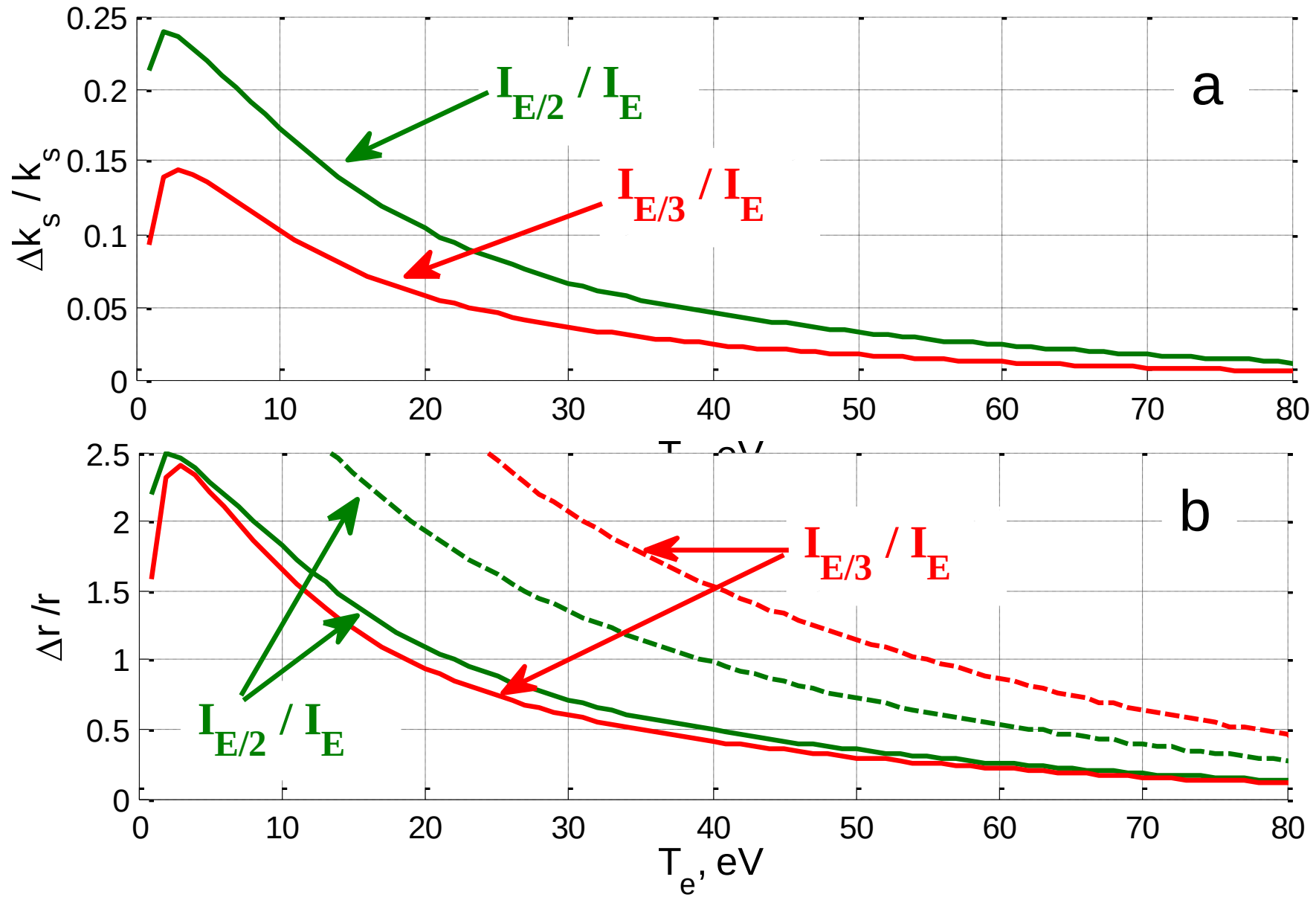

**Figure 8.** ) The required accuracy of the model parameters measurements for evaluation of electron temperature with accuracy 10%; a) – attenuation coefficients relative measurement accuracy, b) relative accuracy of measurement of the beam transverse dimension $r_b$ (solid line) and radius of plasma (dash line)

In experiments on the GOL-NB facility, beam attenuation is measured using a secondary emission probe located on the axis of the neutral beam in the beam dump tank (figure 1). A profilometer of the beam passing through the plasma can be used to determine the radial profile of the plasma [15]. This same profilometer can be used to determine the profile of the beam current density when measuring in the absence of plasma. The dependence of plasma density on radius and beam current density on distance from the axis can be approximated by Gaussian functions:

$$\frac{1}{\sqrt{2\pi}}\frac{\overline{nl}}{r_{pl}}exp\left(-\frac{r^2}{2r_{pl}^2}\right)$$

$$J(x) = J_0 exp\left(-\frac{x^2}{2r_b^2}\right)$$

where $r_{pl}$, $r_b$ – are the characteristic diameter of the plasma and beam transverse size measured by the profilometer, and $\overline{nl}$ – is the linear density of the plasma on the central chord.

The linear plasma density can be found from the attenuation of the neutral beam on the central chord using the equation:

$$k_s = \frac{J_{att}}{J_0} = \frac{\sum_E J_0^E \, exp\left(-\overline{nl}\cdot\sigma_{loss}^E(T_e)\right)}{\sum_E J_0^E}$$

where $J_{att}/J_0$ is the measured attenuation of the neutral beam by the secondary emission probe, $J_0^E$ is the equivalent current of the beam fraction with energy E measured in experiments with beam injection into gas, and $\sigma_{loss}^E$ is the effective capture cross-section of fast atoms with energy

E. It should be noted that the effective capture cross-section depends on the electron temperature of the plasma, so the concentration n(r) can only be found assuming a certain plasma temperature.

The intensity of the measured Doppler-shifted Hα line for the beam fraction with energy E observed at an angle φ is:

$$I^E = \langle \sigma_{H\alpha}^E v^E \rangle \int_{-x0}^{x0} dx \cdot \frac{J(x)}{e} \cdot k_{deex}^E \left( n\left( \frac{x}{sin(\varphi)} \right) \right) \cdot n\left( \frac{x}{sin(\varphi)} \right) \cdot k_{att}^E(n, T_e, x)$$

where $\langle \sigma_{H\alpha}^E v^E \rangle$ – $H_\alpha$ emissivity rate (see section 3.2), $k_{deex}^E$ – is the collisional de-excitation coefficient, (see section 3.4), $k_{att}^E$ is the attenuation factor, which determines the attenuation of the beam in the plasma before reaching the observation point, ±x0 – are the limits of radiation detection over chord of registration, and x corresponds to the direction across the plasma axis and the beam axis, while z corresponds to the direction along the beam axis.

The attenuation coefficient $k_{att}^E$ can be found by integrating over the path of the beam:

$$k_{att}^E(n, T_e, x) = exp\left( -\sigma_{loss}^E(T_e) \int_{-z0}^{x \cdot ctg(\varphi)} dz \cdot n\left( \sqrt{z^2 + x^2} \right) \right)$$

The plasma temperature can be found by comparing the calculated and measured ratios of the emission intensities of different beam fractions (see section 2).

The accuracy of the model parameter measurements required to measure the temperature with a given accuracy can be estimated as follows.

$$\left( \frac{\Delta p}{p} \right)^2 = \left( \frac{\Delta T_e}{T_e} \cdot \frac{T_e}{p} \cdot \frac{\partial \left( I^{E_frac}/I^{E0} \right)}{\partial T_e} \cdot \frac{\partial p}{\partial \left( I^{E_frac}/I^{E0} \right)} \right)^2$$

where Δp/p is the relative measurement error of parameter p, and ΔTe/Te is the required accuracy of temperature measurement. Figure 8 shows the measurement accuracies of model parameters (attenuation coefficient on the central chord $k_s$, characteristic beam transverse dimension $r_b$, and plasma radius $r_{pl}$) required to determine the plasma temperature with an accuracy of 10%, calculated for typical experimental parameters on GOL-NB ($k_s$=0.5, $r_b$=4 cm, $r_{pl}$=4.5 cm). As can be seen from the figure, measuring the beam attenuation with an accuracy of a few percent is necessary for temperature measurement. The size of the plasma and beam radius has a weak influence on determining the plasma temperature.

## 5. Conclusion

The proposed method allows for measuring the electron temperature of plasma up to 40 eV, which is of great importance in the current stage of experiments on the GOL-NB facility. However, temperature measurements in the range above 20 eV require the measurement of line intensities and beam attenuation with a percentage accuracy, which is a challenging task. Moreover, the interpretation of measurement results depends on the reliability of the atomic data used. Therefore, the most promising approach is to use the proposed method together with Thomson scattering temperature measurements. The Thomson scattering system, which is currently under construction, will allow for measuring the electron temperature with high accuracy at a single moment during a shot, enabling calibration of the proposed measurement system and using it to measure temperature dynamics.

## Data availability

The data presented in the figures of this article are available in the form of MATLAB figures at Mendeley Data service DOI: 10.17632/xch3djpnpt.1
The author will be happy to provide data of interest in other formats upon request.

## References

[1] V.V. Postupaev et al. *Start of experiments in the design configuration of the GOL-NB multiple-mirror trap*, *Nuclear Fusion* **62 (2022)** 086003.

[2] V.I. Batkin et al., *Neutral beam injectors for the GOL-NB facility*, *AIP conference proceedings* **1771** (2016) 030010.

[3] V.V. Postupaev and D.V. Yurov *Modeling of reference operating scenario of GOL-NB multiple-mirror trap*, *Plasma Physics Reports* **42** (2016) *1013*

[4] E.N. Sidorov et al, *Four-electrode probe for plasma studies in the gol-nb multiple-mirror trap*, 2021 *JINST* **16** T11006.

[5] A. Ivanov et al., *Characterization of ion species mix of the TEXTOR diagnostic hydrogen beam injector with a rf and arc-discharge plasma box*, *Rev. Sci. Instrum.* **75** (2004) 1822.

[6] R. Uhlemann, R.S. Hemsworth, G. Wang and H. Euringer, *Hydrogen and deuterium ion species mix and injected neutral beam power fractions of the TEXTOR–PINIs for 20–60 kV determined by Doppler shift spectroscopy*, *Rev. Sci. Instrum.* **64** (1993) 974.

[7] R.K. Janev and J.J. Smith *Cross sections for collision processes of hydrogen atoms with electrons, protons and multiply charged ions*, Atomic And Plasma-Material Interaction Data For Fusion, V.4 AEA-APID-4 (1993)

[8] C.F.Barnett *Collisions of H, $H_2$, He and Li atoms and ions with atoms and molecules*, Atomic Data for Fusion, V.1 ORNL-6086/VI report (1990)

[9] https://www-amdis.iaea.org/ALADDIN/

[10] H.P. Summers, (2004) *The ADAS User Manual, version 2.6*, http://www.adas.ac.uk'

[11] http://open.adas.ac.uk

[12] S. Polosatkin, *Effect of sublevel population mixing on the interpretation of doppler-shift spectroscopy measurements of neutral beam content*, 2013 *JINST* **8** P05007 [arXiv:1208.4162]

[13] S. V. Polosatkin; A. A. Ivanov; A. A. Listopad; I. V. Shikhovtsev, *Study of sublevel population mixing effects in hydrogen neutral beams*, *Rev. Sci Instrum.* **85** (2014) 02A707 [arXiv:1309.1339]

[14] I. Bray, A.T. Stelbovics, *Calculation of electrons scattering on hydrogenic targets*, *Advances in Atomic, Molecular and Optical Physics* **35** (1995) pp.209-254

[15] A.V.Nikishin et al., *Multi-chord beam diagnostics of plasma at the GOL-NB device*, *Plasma Physics Reports* **48** (2022) 220